\documentclass[superscriptaddress,preprintnumbers,nofootinbib]{revtex4}
\usepackage{graphicx}

\newcommand{\eV}{\text{eV}}

\begin{document}
\preprint{hep-ph/0606013}
\preprint{KEK-TH-1090}

\title{
Will MINOS see new physics?
}

\author{Noriaki~Kitazawa}
\email{E-mail: kitazawa_at_phys.metro-u.ac.jp}
\affiliation{Department of Physics, Tokyo Metropolitan University,
Hachioji, Tokyo 192-0397, Japan}

\author{Hiroaki~Sugiyama}
\email{E-mail: hiroaki_at_post.kek.jp}
\affiliation{Theory Group, KEK, Tsukuba, Ibaraki 305-0801, Japan}

\author{Osamu~Yasuda}
\email{E-mail: yasuda_at_phys.metro-u.ac.jp}
\affiliation{Department of Physics, Tokyo Metropolitan University,
Hachioji, Tokyo 192-0397, Japan}

\begin{abstract}
The effect of non-standard neutrino interactions with matter
at long baseline neutrino experiments is examined
in a model independent way, taking into account the constraints
from all the experiments.
It is found that such a non-standard interaction can
enhance the flavor transition probability so significantly
that the ongoing experiment MINOS may see the signal
of $\nu_\mu\rightarrow\nu_e$ which is much more than the
prediction by the standard  scenario.
It is also found that the silver channel $\nu_e\to\nu_\tau$ at
a neutrino factory could have a huge enhancement.
\end{abstract}


\maketitle

It has been shown from the measurements
of the atmospheric, solar, reactor, and accelerator neutrinos,
that neutrinos have masses and mixings~\cite{Eidelman:2004wy}.
Furthermore it is expected that precise measurements of the
neutrino oscillation parameters will be performed at
future long baseline neutrino oscillation
experiments with intense neutrino beams, such as
T2K~\cite{Itow:2001ee}, a next generation experiment
which is now under construction, and
a neutrino factory~\cite{Geer:1997iz},
a beta-beam experiment~\cite{Zucchelli:2002sa}, etc.
Just like the goal of the B physics~\cite{Eidelman:2004wy} has shifted from
measuring the mixing angle and the CP phase in the standard model
to probing the new physics effects which would show up as deviation
from the standard model, there have been serious discussions
on search for new physics at
these long baseline experiments in the future~\cite{ISS-physics}.

In the past a class of non-standard neutrino
interactions with matter
were considered to explain the flavor transitions of the
solar~\cite{solar-NSI} or atmospheric
neutrinos~\cite{Gonzalez-Garcia:1998hj} without the standard
oscillations due to masses.  While the scenarios without the standard
oscillations have been disfavored, the non-standard interactions
themselves can still exist as perturbation to the standard
oscillations.  This class of the non-standard interactions offers us
an interesting possibility which can be tested at the future long baseline
experiments because existence of such interactions
would give us some clue about the new
physics beyond the standard model.

In this Letter,
we investigate the effect of the non-standard interaction with matter
upon the oscillation probabilities in long baseline experiments
by taking into account the
constraints in~\cite{Davidson:2003ha,Friedland:2005vy},
which showed that some terms of the non-standard interactions can have
size of $O(1)$ relative to the standard interaction with
matter.

Here we consider the following four-fermi interactions:
\begin{eqnarray}
{\cal L}_{\text{eff}}^{\text{NSI}} =
-2\sqrt{2}\, \epsilon_{\alpha\beta}^{fP} G_F
(\overline{\nu}_\alpha \gamma_\mu P_L \nu_\beta)\,
(\overline{f} \gamma^\mu P f),
\label{NSIop}
\end{eqnarray}
where only the interactions with $f = e, u, d$ are relevant to
the flavor transition of neutrino due to the matter effect,
$G_F$ denotes the Fermi coupling constant, $P$ stands for
a projection operator and is either
$P_L\equiv (1-\gamma_5)/2$ or $P_R\equiv (1+\gamma_5)/2$.
(\ref{NSIop}) is the most general form of the
interactions which conserve electric charge, color, and
lepton number~\cite{Davidson:2003ha}.
(\ref{NSIop}) is supposed to arise from certain new physics,
but we do not specify any particular dynamics which produces
(\ref{NSIop}), so our approach is model-independent
in this sense~\footnote{
One way to escape from the strong constraints on
$\epsilon_{\alpha\beta}$
by the experiments with the charged lepton
is to argue that the operator in (\ref{NSIop}) originates
from the dimension-eight one such as
$\left( \overline{L}_\alpha P_R H^c \right) \gamma_\mu
\left( (H^c)^\dagger P_L L_\beta\right)
\overline{f} \gamma^\mu P f$,
where $H$ and $L$ denote $SU(2)_L$ doublet
of the higgs and lepton, respectively.
After the breaking of $SU(2)_L$
with the vacuum expectation value $v$ of the neutral higgs,
this operator results in
$v^2 ( \overline{\nu}_\alpha \gamma_\mu P_L \nu_\beta )
( \overline{f} \gamma^\mu P f )$
which is nothing but the operator in (\ref{NSIop}).
Under the assumption,
$\epsilon_{\alpha\beta}^{fP}$
is independent of the coupling constants for
$( \overline{l}_\alpha \gamma_\mu P_L l_\beta )
( \overline{f} \gamma^\mu P f )$
from the model-independent viewpoint
because (\ref{NSIop}) is obtained
after the $SU(2)_L$ breaking.
Then,
the constraint on $\epsilon_{\alpha\beta}^{fP}$,
can be obtained only by the experiment with neutrinos.
}.
There can be also interactions that predict flavor transitions
at production or detection of neutrinos~\cite{Grossman:1995wx},
but here we will discuss
only the effect in propagation of neutrinos for simplicity.
Many people~\cite{LBL-NSI,Huber:2002bi} have discussed
the effects of the non-standard interactions in propagation
of neutrinos at future long baseline oscillation experiments,
but the present work is the first to consider potentially
large effects of
the new physics terms, which are comparable in magnitude to the standard
matter term, at the long baseline experiments.

These operators introduce
a new potential for the neutrino propagation in the matter,
and the evolution equation in the flavor basis
($\alpha, \beta = e, \mu, \tau$) is given by%
~\footnote{
Throughout this Letter,
we assume that the number of light neutrinos is three
and there is no unitarity violation.
}
\begin{eqnarray}
\hspace{-45mm}
i\frac{d}{dt} \left(
\begin{array}{c}
\nu_e\\
\nu_\mu\\
\nu_\tau
\end{array}
\right)
= {\cal H}
\left(
\begin{array}{c}
\nu_e\\
\nu_\mu\\
\nu_\tau\\
\end{array}
\right), \nonumber \\
{\cal H} =
U
\left(
\begin{array}{ccc}
0 & 0 & 0\\
0 & \frac{\Delta m^2_{21}}{2E} & 0\\
0 & 0 & \frac{\Delta m^2_{31}}{2E}
\end{array}
\right)
U^\dagger
+A\left(
\begin{array}{ccc}
1+ \epsilon_{ee} & \epsilon_{e\mu} & \epsilon_{e\tau}\\
\epsilon_{e\mu}^\ast & \epsilon_{\mu\mu} & \epsilon_{\mu\tau}\\
\epsilon_{e\tau}^\ast & \epsilon_{\mu\tau}^\ast & \epsilon_{\tau\tau}
\end{array}
\right),
\label{3f-NSImass}
\end{eqnarray}
where $\Delta m^2_{jk}\equiv m^2_j-m^2_k$ is the mass squared
difference, $E$ is the neutrino energy,
$A \equiv \sqrt{2} G_F n_e
\simeq
1.0\times 10^{-13} \eV
\left( \rho/2.7\text{g}\cdot\text{cm}^{-3} \right)$
stands for the magnitude of
the standard matter effect,
$n_e$ is the number density of the electron in the matter,
$\rho$ stands for the matter density,
and $U$ is the Maki-Nakagawa-Sakata matrix
in the standard parametrization~\cite{Eidelman:2004wy}.
$\epsilon_{\alpha\beta}$ are defined as
$\epsilon_{\alpha\beta}
\equiv \sum_{f,P} \frac{n_f}{n_e} \epsilon_{\alpha\beta}^{fP}
\simeq \sum_{P}
\left(
\epsilon_{\alpha\beta}^{eP}
+ 3 \epsilon_{\alpha\beta}^{uP}
+ 3 \epsilon_{\alpha\beta}^{dP}
\right)$,
where $n_f$ is the number density of $f$ in matter,
and we have taken into account the fact that the number density of
$u$ quarks and $d$ quarks are three times as that of
electrons.
The striking feature of (\ref{3f-NSImass}) is that
the flavor transition is possible
even at high energy
because the last term in (\ref{3f-NSImass}) is not diagonal,
while the transition vanishes at high energy in the standard case.

$\epsilon_{\alpha\beta}^{fP}$ is a dimensionless parameter
normalized by $G_F$, and theoretically it is expected that
$\epsilon_{\alpha\beta}^{fP}$ is suppressed by a factor
(W boson mass)$^2$/(new physics scale)$^2$.
Experimentally, however, it is known~\cite{Davidson:2003ha}
that some of $\epsilon_{\alpha\beta}^{fP}$ have a very weak bound.
The result in \cite{Davidson:2003ha} is given by
\begin{eqnarray}
\left(
\begin{array}{ccc}
-4 < \epsilon_{ee} < 2.6 & |\epsilon_{e\mu}| < 3.8\times 10^{-4}
& |\epsilon_{e\tau}| < 1.9\\
& -0.05 < \epsilon_{\mu\mu} < 0.08
& |\epsilon_{\mu\tau}| < 0.25\\
& & |\epsilon_{\tau\tau}| < 18.6
\end{array}
\right).
\label{b-eps-0}
\end{eqnarray}

Furthermore, it was shown~\cite{Friedland:2005vy} that the measurements
of the atmospheric and accelerator neutrinos
give non-trivial constraints on $\epsilon_{ee}$,
$\epsilon_{e\tau}$ and $\epsilon_{\tau\tau}$.
It was found in~\cite{Friedland:2005vy} that a strong
constraint applies to the channel $\nu_\mu\to\nu_\mu$
in the high energy atmospheric neutrino data
while there is some freedom
left in the channel $\nu_e\leftrightarrow\nu_\tau$
because neither electron nor tau events are observed
at high energy.
From Fig.6 of~\cite{Friedland:2005vy},
we can read off the following two approximate constraints:
\begin{eqnarray}
&& |\epsilon_{e\tau}| \lesssim |1 + \epsilon_{ee}|,\label{b-eps-2}
\label{b-eps-23}\\
&& \epsilon_{e\tau}^2
\simeq \epsilon_{\tau\tau} \left( 1 + \epsilon_{ee} \right)
\label{b-eps-3}.
\end{eqnarray}
(\ref{b-eps-3}) is the condition for which the survival
probability $P(\nu_\mu\to\nu_\mu)$ of the high energy
atmospheric neutrinos in the presence of the new physics
is reduced to that in the standard case.

Thus, combining (\ref{b-eps-0}), (\ref{b-eps-23}) and (\ref{b-eps-3}),
the region for $\epsilon_{\alpha\beta}$ that we will use
in the following analysis can be summarized as
\begin{eqnarray}
\left(
\begin{array}{ccc}
-4 < \epsilon_{ee} < 2.6 & \epsilon_{e\mu} = 0
& |\epsilon_{e\tau}| < 1.9\\
& \epsilon_{\mu\mu} = 0
& \epsilon_{\mu\tau} = 0\\
& & |\epsilon_{\tau\tau}| < 1.9
\end{array}
\right)
\label{b-eps}
\end{eqnarray}
together with (\ref{b-eps-23}) and (\ref{b-eps-3}).
For simplicity we will not discuss the complex phases of $\epsilon_{\alpha\beta}$
in the following.

Before discussing the three flavor case, it is instructive
to consider the two-flavor scenario because we can express
the oscillation probability analytically.
The Hamiltonian for this case is
\begin{eqnarray}
U
\left(
\begin{array}{cc}
0 & 0\\
0 & \frac{\Delta m^2}{2E}
\end{array}
\right)
U^\dagger
+
A
\left(
\begin{array}{cc}
1+ \epsilon_{ee} & \epsilon_{e\tau}\\
\epsilon_{e\tau} & \epsilon_{\tau\tau}
\end{array}
\right).
\label{2f-NSImass}
\end{eqnarray}
It is easy to diagonalize the matrix,
and we obtain the squared-mass difference $\Delta m^2_M$,
the mixing $\theta_M$ and the oscillation probability
$P(\nu_e\to\nu_\tau)$ at distance $L$ in matter:
\begin{eqnarray}
\left( \frac{\Delta m^2_ML}{4E} \right)^2
&=&
\left( \frac{\Delta m^2L}{4E} \cos{2\theta}
- \frac{AL}{2} (1+\epsilon_{ee}-\epsilon_{\tau\tau}) \right)^2
\nonumber\\
&+& \left( \frac{\Delta m^2L}{4E} \sin{2\theta}
+ AL\epsilon_{e\tau} \right)^2,
\label{2f-mass}\\
\tan{2\theta_M}
&=&
\frac{ \Delta m^2/2E \sin{2\theta} + 2A \epsilon_{e\tau} }
{ \Delta m^2/2E \cos{2\theta} - A (1+\epsilon_{ee}-\epsilon_{\tau\tau})},
\label{2f-ang}\\
P(\nu_e\to\nu_\tau)&=&\sin^2{2\theta_M}\sin^2\left(\frac{\Delta m^2_ML}{4E}\right).
\label{2f-pro}
\end{eqnarray}
To have a large value of the oscillation probability
$P(\nu_e\to\nu_\tau)$, we need large values for both $\sin^2{2\theta_M}$
and $\sin^2\left(\Delta m^2_ML/4E\right)$.
From (\ref{2f-pro}) we observe the two things.
First of all,
the effect of the new physics
in $\sin^2\left(\Delta m^2_ML/4E\right)$ appears
in a form $AL(\epsilon_{ee}-\epsilon_{\tau\tau})$ or
$AL\epsilon_{e\tau}$, so large deviation of $\Delta m^2_ML/4E$
from the standard one $\Delta m^2L/4E$ requires
$AL\epsilon_{\alpha\beta}$ be nonnegligible irrespective of the neutrino
energy $E$.
Secondly, for the experiments with $|\Delta m^2|L/E\simeq {\cal O}(1)$,
we see by multiplying $L$ both the numerator and the denominator of (\ref{2f-ang}) that
the condition for nontrivial contribution of the new physics to
the mixing angle $\theta_M$ again demands that
$AL\epsilon_{\alpha\beta}$ be nonnegligible.
These imply that the baseline length has to
be relatively large for the new physics effect to affect
both of the factors in the oscillation probability,
since $A$ can be roughly estimated as
$A\simeq$1/(2000km) with $\rho\simeq$3g/cm$^3$.
Although it is difficult to treat the three flavor case analytically,
these features hold also in the case with three flavors,
and they are important to
understand the sensitivity of the long baseline experiments
to the new physics.
Typical ongoing and future long baseline experiments and their
baseline length $L$ are:
a reactor experiment~\cite{Anderson:2004pk} ($L\sim$2km),
the T2K experiment~\cite{Itow:2001ee} ($L=$295km),
the MINOS experiment~\cite{Tagg:2006sx} ($L=$735km),
the NOvA experiment~\cite{Ayres:2004js} ($L\simeq$800km),
the T2KK experiment~\cite{T2KK} ($L\sim$1000km),
a neutrino factory~\cite{Geer:1997iz} ($L\sim$3000km).
All these experiments are designed mainly to probe neutrino oscillations
with the atmospheric neutrino mass squared difference
$|\Delta m^2_{\text{atm}}|\simeq 2.5\times10^{-3}$eV$^2$
and the typical neutrino energy $E$ of each experiment satisfies
$|\Delta m^2_{\text{atm}}|L/E\simeq {\cal O}(1)$.
The baseline lengths $L$ of these experiments, however,
are quite different, and
when $\epsilon_{\alpha\beta}\sim{\cal O}(1)$,
only the experiments with nonnegligible value of
$AL$ have sensitivity to the new physics.
A reactor experiment satisfies $AL\ll 1$, so that it has no hope to see
the signal due to $\epsilon_{\alpha\beta}$.  On the other hand,
a reactor experiment has advantage of having no backgrounds
due to the new physics
in measurements of the standard oscillation parameters.
For the T2K experiment, $AL\simeq 3/20$, so it has potential to
see the new physics effect.
For MINOS, NOvA, T2KK and a neutrino factory, since $AL$ is larger,
they have in principle even more potential to see the signal
of $\epsilon_{\alpha\beta}$.

\begin{table}[t]
\tabcolsep = 1mm
\begin{tabular}{|c||c|c|c|c|c|c|c|c|c|}
\hline
set & (a) & (b) & (c) & (d) & (e) & (f) & (g) & (h) & (i)\\
\hline\hline
$\epsilon_{ee}$ & $-4$ & $-4$ & $-4$ & 0 & 0 & 0 & 2.6 & 2.6 & 2.6\\
\hline
$\epsilon_{e\tau}$ & $-1.9$ & 0 & 1.9 & $-1$ & 0 & 1 & $-1.9$ & 0 & 1.9\\
\hline
$\epsilon_{\tau\tau}$ & $-1.2$ & 0 & $-1.2$ & 1 & 0 & 1 & 1 & 0 & 1\\
\hline
\end{tabular}
\caption{
The table shows nine sets of
the reference values of $\epsilon_{\alpha\beta}$ used in the figures.
These sets are consistent with the bounds
(\ref{b-eps-2}), (\ref{b-eps-3}), and (\ref{b-eps}).
}
\label{tab:eps}
\end{table}

Let us now investigate the three flavor case and
see how much the non-standard matter effect
with $\epsilon_{\alpha\beta} = O(1)$
changes the oscillation probabilities in long baseline experiments.
Since the analytical treatment of the three flavor case is difficult,
we will evaluate the oscillation probability numerically.
In the present analysis we use the following values of the
parameters:
$\rho = 2.7 \text{g}\cdot\text{cm}^{-3}$,
$Y_e = 0.5$,
$0 < \Delta m^2_{31} = 2.5\times 10^{-3}\eV^2$,
$\Delta m^2_{21} = 8\times 10^{-5}\eV^2$,
$\sin^2{2\theta_{23}} = 1, \ \ \sin^2{2\theta_{12}} = 0.8$,
$\sin^2{2\theta_{13}} < 0.16, \ \ \delta = 0$.
The reference values of $\epsilon_{\alpha\beta}$ that will be used are
given in Table~\ref{tab:eps}.
As is explained in~\cite{Friedland:2005vy},
the best fit values for $\sin{2\theta_{23}}$ and $\Delta m^2_{31}$
for the atmospheric neutrino measurement
are changed by the nonzero values of $\epsilon_{\alpha\beta}$,
and this effect will be taken into account in the following analysis.

\begin{figure}[t]
\hglue -3.0cm
\rotatebox{-90}{\includegraphics[scale=0.65]{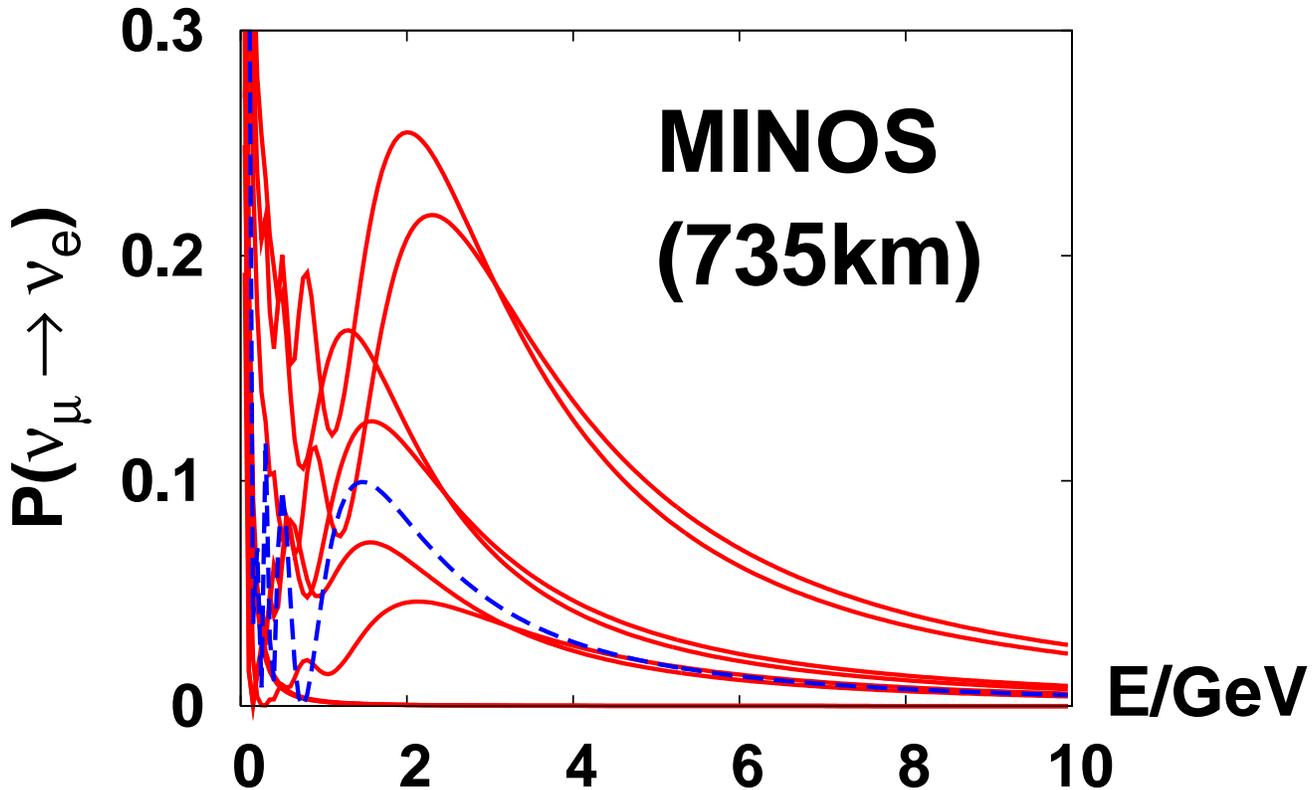}}
\vglue -0.3cm
\caption{
The figure shows $P(\nu_\mu\to\nu_e)$ in MINOS experiment
with the non-standard matter effect.
Solid lines are obtained with the non-standard effect
for $\sin^2{2\theta_{13}} = 0$.
A dashed line is for $\sin^2{2\theta_{13}} = 0.16$
without the non-standard effect.
The set (g) of Table~\ref{tab:eps}
gives the line whose value is the largest
among the nine solid lines at the peak near $E\simeq$2GeV\@.
The sets (b), (e), and (h) give almost the same lines
which vanish for $E\gtrsim$2GeV, where (e) is the case with
$\theta_{13}=0$ and $\epsilon_{\alpha\beta}=0$.
}
\label{fig:MINOS}
\end{figure}

The ongoing MINOS experiment~\cite{Tagg:2006sx}
has the baseline length $L=735$km and the $\nu_\mu$ beam
has a peak at several GeV\@.
The appearance probability $P(\nu_\mu \to \nu_e)$
at MINOS is shown in Fig.~\ref{fig:MINOS}.
We see that the non-standard matter effect due to $\epsilon_{\alpha\beta}$
can give much larger probability
than that by the standard oscillation
with allowed value of $\theta_{13}$.
The enhancement depends mainly on $\epsilon_{e\tau}$,
$\sin{2\theta_{23}}$, and $\Delta m^2_{31}$,
and it almost disappears
if one of these parameters is set to be zero.
Thus,
the nonstandard $\nu_\mu$-$\nu_e$ oscillation
can be understood very roughly
by the standard $\nu_\mu$-$\nu_\tau $ oscillation
and the subsequent $\nu_\tau$-$\nu_e$ transition
with $\epsilon_{e\tau}$.
The sets (b), (e), and (h) in Table~\ref{tab:eps},
which have $\epsilon_{e\tau} = 0$,
give the almost same line that vanishes for $E\gtrsim$2GeV\@.
The sets (a), (c), (g), and (i),
which have large $|\epsilon_{e\tau}|$,
give large values of the probability.
If the MINOS experiment observes the probability $P(\nu_\mu \to \nu_e)$
that is given by the set (g) in Table~\ref{tab:eps}, which is the
most extreme case,
then it would be a clear signal of the existence
of the non-standard interaction,
because the signal is much
larger than the standard prediction with the maximum possible value of
$\theta_{13}$,
even if we take into consideration the 30\% uncertainty in
$\sin^2\theta_{23}$.~\footnote{If the value of $\theta_{13}$
is close to the current bound given by the CHOOZ
experiment~\cite{Apollonio:2002gd}, then the oscillation
probability for $\nu_\mu \to \nu_e$ is proportional to
$\sin^2\theta_{23}\sin^22\theta_{13}$.
}
On the other hand, even if MINOS does not see any signal
of $\nu_\mu \to \nu_e$ within its sensitivity,
it would still give us new information on the allowed region of
$\epsilon_{\alpha\beta}$, which will be reported
elsewhere~\cite{ksy2}.
NOvA experiment~\cite{Ayres:2004js} at $L\simeq800\text{km}$
will have better sensitivity to the non-standard interaction
because of fewer backgrounds with the off-axis beam.

\begin{figure}[t]
\hglue -3.0cm
\rotatebox{-90}{\includegraphics[scale=0.65]{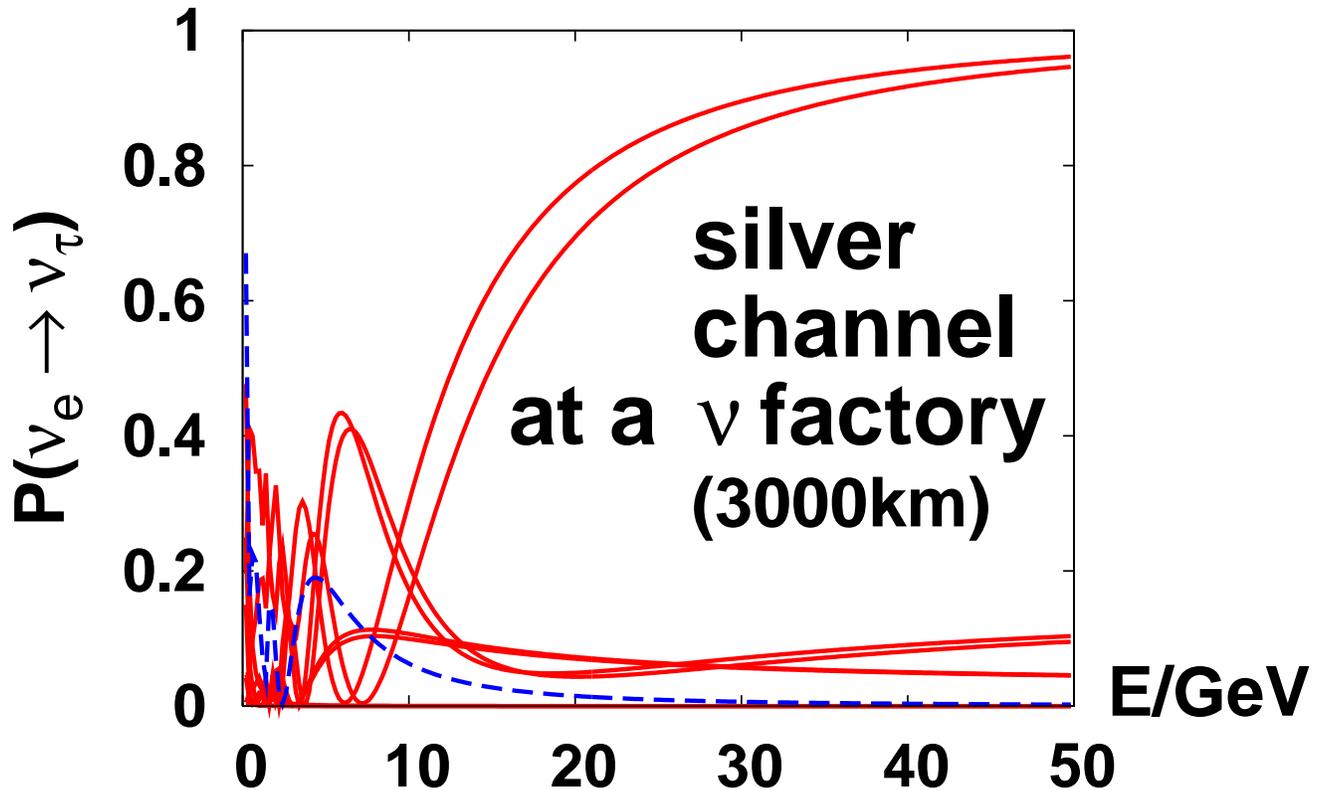}}
\vglue -0.3cm
\caption{
The figure shows $P(\nu_e\to\nu_\tau)$ in the neutrino factory
with the non-standard matter effect.
Solid lines are obtained with the non-standard effect
for $\sin^2{2\theta_{13}} = 0$.
A dashed line is for $\sin^2{2\theta_{13}} = 0.16$
without the non-standard effect.
The sets (d) and (f) of Table~\ref{tab:eps}
give the line that is $\simeq 1$ at high energy.
The sets (b), (e), and (h) give approximately the same lines
that have almost no effect of the non-standard interaction.
}
\label{fig:silver}
\end{figure}

We have also considered the case of a neutrino factory~\cite{Geer:1997iz}
with a baseline length $L=3000\text{km}$ and the muon energy
$E_\mu=50$GeV and the neutrino beam has a peak at around 30GeV\@.
For the so-called golden channel $P(\nu_e \to \nu_\mu)$~\cite{Cervera:2000kp},
the qualitative behavior is similar to that in MINOS\@;
There is a large enhancement of the probability
for neutrinos of a few GeV\@.
On the other hand, for
the so-called silver channel $P(\nu_e \to \nu_\tau)$~\cite{Donini:2002rm}
the probability
has a huge enhancement at high energy
as is shown in Fig.~\ref{fig:silver},
and the probability becomes independent of the energy
in the region.
It is because
the terms of $|\Delta m^2_{jk}|/E$ in (\ref{3f-NSImass})
disappear at high energy
and $\epsilon_{\alpha\beta}$ gives
the energy independent transition of flavors.
In fact, the oscillation probability
$P(\nu_e \to \nu_\tau)$ at high energy is approximately given by
the two flavor formula (\ref{2f-pro}),
since $\nu_\mu$ decouples from $\nu_e$ or $\nu_\tau$
with the Hamiltonian (\ref{3f-NSImass}) at high energy.

In this Letter,
we discussed the effects of the non-standard interaction with matter
upon the long baseline oscillation experiments, with
only the terms whose sizes have a very weak bound
from other neutrino experiments.  Taking into
account all the constraints on the non-standard
interactions including the atmospheric neutrino data,
the 
$ee$, $e\tau$, $\tau\tau$ components of the non-standard matter effect
can be comparable in magnitude to
the standard matter interaction.
Possibility of detecting
such a potentially large effect was examined
for the first time in
the ongoing and future long baseline experiments.
It was found that $\epsilon_{\alpha\beta}$
can give a large oscillation probability
$P(\nu_\mu\to\nu_e)$ in the ongoing MINOS experiment.
In the most optimistic case,
the oscillation probability is so large that it cannot be explained
by the standard oscillation with $\theta_{13}$.
Thus,
it is a clear signal of the non-nonstandard interaction
if MINOS observes such a large number of $\nu_e$ appearance events.
It was also shown that the oscillation probability $P(\nu_e\to\nu_\tau)$
of the silver channel at a neutrino factory
can be very large at high energy
due to $\epsilon_{\alpha\beta}$ of $O(1)$;
Amazingly,
even $P(\nu_e\to\nu_\tau) \simeq 1$ is possible.
Therefore,
the silver channel is a very promising
way to look for the effect of the non-standard interaction.
Once such signals are found, it would be necessary to
separate the effects due to the standard oscillations
and those due to the new physics~\cite{Huber:2002bi}.
In that case, as was explained in the text,
combination with a reactor experiment or long baseline
experiments with different baseline lengths would be important.

In conclusion, there remains a possibility
to find the large effect of the non-standard interaction
in the long baseline oscillation experiments.
More detailed analyses as well as other
discussions on the subject will be given elsewhere~\cite{ksy2}.

We thank the participants
of the 3rd Plenary Meeting of the International Scoping Study
of a Future Neutrino Factory and Superbeam Facility~\cite{ISS},
where the contents of the present work were presented,
for valuable discussions.
This work was supported in part by Grants-in-Aid for Scientific Research
No.\ 16340078, Japan Ministry
of Education, Culture, Sports, Science, and Technology,
and the Research Fellowship
of Japan Society for the Promotion of Science
for young scientists.


\end{document}